\documentclass[a4paper,12pt]{article}
\pdfoutput=1

 \usepackage[latin1]{inputenc}
\usepackage[T1]{fontenc}
\usepackage{amsmath}
\usepackage{amsfonts}
\usepackage{amssymb}
\usepackage{color}
\usepackage{graphicx}
\usepackage{caption}
\usepackage{comment}
\usepackage{subcaption}
\usepackage{hyperref}
\hypersetup{colorlinks=true,linkcolor=blue,citecolor=blue} 
 \newcommand{\be}{\begin{equation}}
	 \newcommand{\ee}{\end{equation}}
	 \newcommand{\ba}{\begin{eqnarray}}
		 \newcommand{\ea}{\end{eqnarray}}
		 
		   \newcommand{\bea}{\begin{eqnarray}}
			 \newcommand{\eea}{\end{eqnarray}}
 \newcommand{\nn}{\nonumber}

\newcommand{\nl}{\nonumber \\} 
\newcommand{\beq}{\begin{equation}}
\newcommand{\eeq}{\end{equation}}
\newcommand{\p}{\partial}
\def\a{\alpha}

\def\le{\left(}
\def\ri{\right)}
\def\tx{\tilde{x}}

\def\tx{{\tilde{x}}}

\begin{document} 

\title{The holographic vortex lattice using the circular cell method}
\author{Gianni Tallarita$^{a}$ and Roberto Auzzi $^{b,c}$ \\\\
  $^{a}$ {\it \small Departamento de Ciencias, Facultad de Artes Liberales,
 Universidad Adolfo Ib\'a\~nez,}\\ 
 { \it \small  Santiago 7941169, Chile,}\\\\
   $^{b}$ {\it \small  Dipartimento di Matematica e Fisica,  Universit\`a Cattolica
del Sacro Cuore, }\\
{\it \small  Via Musei 41, 25121 Brescia, Italy, } 
\\\\
$^c$ {\it \small INFN Sezione di Perugia,  Via A. Pascoli, 06123 Perugia, Italy.}\\\\
 }
\date{\hfill}

\maketitle
\begin{abstract}

We investigate vortex lattice solutions in a holographic superconductor model
in asymptotically AdS$_4$ spacetime
which includes the gravitational backreaction of the vortex. The circular cell 
approximation, which is known to give a good result for several physical
quantities in the Ginzburg-Landau model, is used.
The critical magnetic fields and the magnetization curve are computed.
The vortex lattice profiles are compared to expectations from the Abrikosov solution
in the regime nearby the upper critical magnetic field $H_{2c}$
for which superconductivity is lost.

\end{abstract}

\section{Introduction}

An important and generic property of higher temperature superconductors
is the presence of a strange metal state found just above
the superconducting critical temperature.
The transport properties of strange metals are very different
from the ones of conventional Fermi liquid.
In particular, the standard quasi-particle picture
does not give a useful description of the physics of the system \cite{Sachdev-book,Hartnoll:2016apf}.
An interesting class of models without a quasi-particles description
can be built using the AdS/CFT correspondence. 
The correspondence maps a strongly interacting quantum system
in the boundary to a classical gravity problem in the bulk, 
and so it provides a controlled environment in which to study 
strongly coupled systems.

Since Abrikosov's seminal work \cite{Abrikosov:1956sx},
the magnetic properties of type II superconductors have 
been the subject of many experimental and theoretical studies
 (see \cite{brandt1} for a review).
In this phase magnetic flux penetrates the superconductor by forming vortices, 
which are arranged in lattice geometries.
Using  several microscopic techniques,
these periodic arrays of vortices have been
experimentally studied in the lab  both for conventional
 and for higher temperature superconductors.
 
 The Ginzburg-Landau (GL) theory is a very useful macroscopic description
 of superconductors (see \cite{Tinkham-book} for a textbook) 
 which can be used to model the Abrikosov
 vortex lattice in a quantitative way. Strictly speaking,
 the GL theory is valid only close to the critical temperature;
 indeed, it can be derived from the Bardeen-Cooper-Schrieffer
 (BCS) theory just in this regime. Given the experimental importance
 of vortex lattices in higher-temperature superconductors, 
 it is important to theoretically study vortex
 lattices also in theoretical situations where no quasi-particle 
 picture is available. Holographic superconductors 
 \cite{Gubser:2008px,Hartnoll:2008vx,Hartnoll:2008kx,Zaanen-book} provide a 
 controlled theoretical laboratory to explore situations
where the quasi-particle approximation is not applicable,
and so they may give precious hints on the behaviour 
 of vortex lattices in non-conventional superconductors. 

Vortices in holographic superconductors have been studied
by many authors.   Most of the early studies 
\cite{Albash:2009iq,Montull:2009fe,Keranen:2009re,
Domenech:2010nf,Tallarita:2010vu,Iqbal:2011bf,Tallarita:2015mca} 
neglect the gravitational backreaction of the vortex solution.
This is a well justified approximation in the regime where
the scalar condensate is small.
A systematic study of gravitational backreaction
in the case of a single vortex was performed in \cite{Dias:2013bwa}. 
 This analysis allowed to systematically compute
 thermodynamic properties of the vortex.

The study of vortex lattices is more complicated because 
 there is no cylindrical symmetry, and so one needs to solve
 a partial differential equation with an extra dynamical 
variable. Without backreaction, a study of the holographic
vortex lattice was initiated in \cite{Maeda:2009vf}. 
Vortex lattices with backreaction were studied in the AdS$_2$ $\times R^2$ geometry
in  \cite{Bao:2013fda,Bao:2013ixa}: this geometry describes the
near horizon limit of extremal magnetic Reissner-Nordstrom black holes,
and so it is relevant for the zero temperature limit of a holographic vortex lattice.
The vortex lattice in a holographic model with 
$SU(2)$ gauge field was studied in \cite{Bu:2012mq}.
Vortex lattices in holographic superfluid were studied in 
\cite{Xia:2019eje,Li:2019swh}.

The problem of constructing the fully backreacted holographic vortex lattice,
valid for all ranges of magnetic fields and temperatures, remains therefore an unsolved problem. 
This is not surprising, the problem involves hard numerical computations, 
with complicated starting ansatze for the metric and matter fields. 
In this paper we will provide an approximate solution to this problem using
the circular cell method (CCM) \cite{ihle}, which is a technique 
already used for vortex lattices in  the
Landau-Ginzburg framework \cite{brandt2,brandt2-bis,brandt3}.

\begin{figure}[ptb]
\centering
\includegraphics[width=0.8\linewidth]{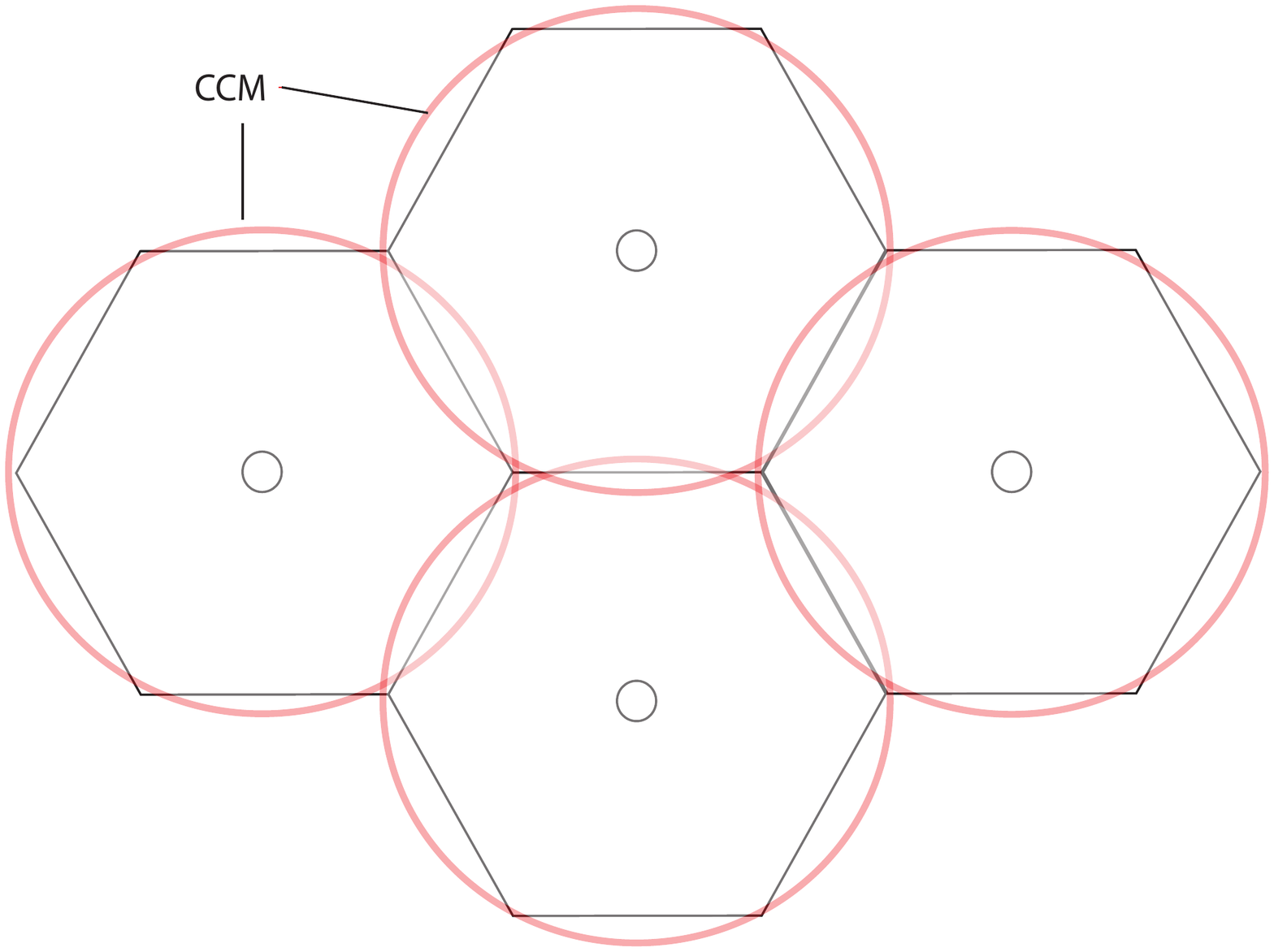}
\caption{Schematic representation of Circular Cell Method}
\label{fig0}
\end{figure}

The CCM approximates the full geometrical lattice solution by replacing 
each cell of the lattice with a circular one of the same area (see Figure \ref{fig0}). 
This dramatically simplifies the problem as one can use a cylindrical 
symmetry ansatz to simplify the calculation. For the case of standard Abrikosov lattices, the method 
is extremely accurate over the whole range of magnetic fields, with several physical 
quantities such as critical magnetic field and magnetization curve
differing by percent level
between the full geometrical result and the CCM \cite{brandt2,brandt2-bis,brandt3}. 
This remarkable  result serves as motivation to use this method 
in the holographic context. We will however quantify the
validity of the approximation, at least in the previously mentioned limit of critical fields, 
where an analytic solution is available. 

The paper is organized as follows: in section \ref{sect:theo-setting} we provide an introduction
 to the theoretical setting with which we will work throughout the paper.
 In section \ref{sect:CCM} we will apply the CCM to the holographic
 vortex lattice and we will compute the magnetization curve of
 the superconductor. In section \ref{sect:critical} we will
  compare the circular cell approximation to the Abrikosov 
  solution in the limit of critical magnetic field $H_{2c}$.
 We conclude in section \ref{sect:conclu}.

 \section{Theoretical setting}
 \label{sect:theo-setting}
 
 We will consider the same model
 of holographic superconductor 
 in asymptotically AdS$_4$ spacetime
  as in \cite{Dias:2013bwa}.
 The bulk lagrangian is:
\be\label{sano}
S_{ANO} = \frac{1}{16\pi G_N}\int d^4x \sqrt{-g}\left[R+\frac{6}{L^2}
-\frac{1}{2}F_{\mu \nu}F^{\mu \nu }-2(D_\mu \phi)(D^\mu \phi)^{\dagger}-2V(|\phi|^2)\right],
\ee
where $L$ is the AdS radius and
\be
\label{pot}
V(|\phi|^2) = -\frac{2}{L^2}|\phi|^2\left(1-\frac{1}{2 }|\phi|^2\right) \, .
\ee
\be
F_{\mu \nu} = \partial_\mu A_\nu -\partial_\nu A_\mu,
\,  \qquad
D_\mu \phi = \partial_\mu \phi - iq A_\mu \phi \, .
\ee
Here $D_\mu$ denotes the combination of the gravity
and $U(1)$ gauge covariant derivatives. 
We use slightly unconventional field normalizations:
in our units  $\phi$ and $A_\mu$ are both dimensionless,
while $q$ has the same dimension as energy
($q L$ is instead dimensionless).
It turns out that the dimensionless quantity $q L$ indirectly characterizes
the ratio between the magnetic penetration and the coherence length:
indeed, as shown in \cite{Dias:2013bwa}, for $q L=1$ the vortex is in the
type II regime while for $q L=2$ it is type I.

The extrema of the potential (\ref{pot}) are
\bea
|\phi| &=& 0  \, , \qquad V=0 \, , \qquad m^2_\phi=-\frac{2}{L^2} \, , \nl
|\phi| &=& 1 \, , \qquad V=-\frac{1}{L^2}  \, , \qquad m^2_\phi=\frac{4}{L^2} \, .
\eea
We will consider the first of these AdS vacua, whose metric is:
\be
ds^2 = \frac{L^2}{z^2}(-dt^2+dz^2+dr^2+r^2d\theta^2) \, .
\label{empty-ads}
\ee
Nearby the boundary, the field $\phi$ has the following expansion
\be
\phi = \alpha z^{\Delta_1}+ \beta z^{\Delta_2} +... \, , \qquad \Delta_1=1 \, , \qquad \Delta_2=2 \, .
\label{bou-phi}
\ee
The  dimensions $\Delta_i$ are the solutions of $m^2_\phi L^2=\Delta (\Delta-3)$. 

 In order to allow for a vortex solution, we first need to 
realize a phase with a non-zero homogeneous scalar 
field outside the vortex core.
In the original holographic superconductor model \cite{Hartnoll:2008vx,Hartnoll:2008kx}
a chemical potential was introduced as a boundary condition of the $A_0$ field,
in order for the scalar field $\phi$ to condense in the bulk.
 It turns out that, in order to achieve the condensation of $\phi$ in a holographic model,
one does not need to consider the complication of non-zero 
background charge density.
As explained in \cite{Faulkner:2010gj}, the condensation of $\phi$
can be achieved also in a neutral black hole background,
 using a Robin boundary condition of the following form:
\be\label{bc}
\beta = \kappa \alpha \, , \qquad \kappa < 0 \, ,
\ee
which is 
 dual to introducing a relevant double-trace deformation
 in the field theory side of the holographic duality
\cite{Witten:2001ua,Berkooz:2002ug}:
 \beq
 \Delta \mathcal{V}=  \kappa \,  \mathcal{O}^\dagger\mathcal{O} \, .
 \eeq
  When $\kappa$ is negative, this term triggers the condensation of the $\mathcal{O}$ operator.
  We choose to perform our calculations in the  double trace deformation 
  because    it avoids the necessity to deal with an extra field profile for $A_0(z)$,
which reduces the number of equations that we have to solve.
   \newline
  
  With zero magnetic field, the critical temperature for condensation \cite{Faulkner:2010gj} is:
  \beq
  T_c=\frac{3}{4 \pi} \frac{\Gamma(1/3)^3}{\Gamma(-1/3) \Gamma(2/3)^2} \kappa \approx -0.62 \kappa \, .
  \label{Tcritica}
  \eeq

The equations of motion resulting from the action in eq.~(\ref{sano}) are:
\beq
\label{eqs}
G_{\mu \nu}=0 \, , \qquad
D_\mu F^{\mu \nu}=i q[(D^\nu \phi)\phi^{\dagger}-(D^\nu \phi)^{\dagger}\phi] =J_\mu \, ,  \qquad
g^{\mu \nu}D_\mu D_\nu \phi-V'(|\phi|^2)\phi=0 \, ,
\eeq
where
\beq
G_{\mu \nu}=
R_{\mu \nu}+\frac{3}{L^2} g_{\mu \nu}-\left[(D_\mu \phi)(D_\nu \phi)^{\dagger}
+(D_\nu \phi)(D_\mu \phi)^{\dagger}+g_{\mu \nu}V(|\phi|^2)+F_\mu^\sigma F_{\sigma \nu }
-\frac{g_{\mu \nu}}{4}F^{\rho \sigma}F_{\rho \sigma }\right] \, .
\eeq

For the dual conformal field theory interpretation,
it is crucial to specify the boundary condition for the $U(1)$
at $z=0$  \cite{Witten:2003ya,Domenech:2010nf}. 
We denote by $\vec{B}$ the four dimensional gauge
curvature tangent to the boundary, so that involves just
$dx_i \wedge dx_j$ terms, where $x_i=(t,r,\theta)$.
The $\vec{E}$ components instead involve 
 terms such as $dx_i \wedge dz$, where $z$ is the AdS$_4$ normal coordinate.
If one chooses Dirichlet boundary condition ($\vec{B}=0$),
the field theory dual is a superfluid; instead with a Neumann boundary condition
 ($\vec{E}=0$) a dynamical gauge field appears in the boundary,
 and the field theory dual is a superconductor.
  The two choices of boundary condition are related by bulk S duality
 \cite{Witten:2003ya}.
The solution that we will discuss in the next section
 has cylindrical symmetry, with angular coordinate $\theta$.
 The only non-zero component of the gauge field is $A_\theta$,
 which has the following expansion nearby the boundary:
 \beq
 A_\theta=a_\theta + z \, J_\theta+ \mathcal{O}(z^2) \, ,
 \label{Atheta}
 \eeq
 where $z$ is the Fefferman-Graham (FG) coordinate 
  for the asymptotically AdS$_4$ backreacted metric.  
 In order to impose the Neumann boundary condition,
we will set $J_\theta=0$ in the boundary condition of the partial differential equation.

\subsection{The normal state}

 In general, the vacuum state of our system with arbitrary temperature and applied magnetic field
must be determined by solving the complicated non-linear system of PDE
in eq. (\ref{eqs}). In the regime above the upper applied critical magnetic
$H_{2c}$ for which superconductivity is lost, the $\phi$ condensate is zero
and the magnetic field $\vec{B}$ is spatially uniform.
The system then is described by the  magnetic Reissner-Nordstrom  (RN) black brane solution:
\beq
ds^2=\frac{L^2}{z^2} \left( -f(z) dt^2 + \frac{dz^2}{f(z)} + d \vec{x}^2 \right) \, ,
\label{Reissner}
\eeq
where 
\beq
f(z)=1-\le 1+ \frac{z_h^4 B^2}{2 L^2}  \ri \le \frac{z}{z_h} \ri^3
+\frac{z_h^4 B^2}{2 L^2} \le \frac{z}{z_h} \ri^4 \, ,
\qquad
A= B \, dx \wedge dy \, ,
\eeq
and $z_h$ is the position of the horizon.
The Hawking temperature $T$ is
\beq
T=\frac{1}{4 \pi z_h} \left( 3  -\frac{z_h^4 B^2}{2 L^2} \right) \, .
\eeq
In order to find $z_h$ as a function of $B,T$, one has to solve a  quartic equation:
\beq
z_h^4+ z_h \le \frac{8 \pi T L^2}{B^2} \ri -\frac{6 L^2}{B^2}=0 \, .
\label{equazione-zh}
\eeq
The solution to eq. (\ref{equazione-zh}) 
can be written in compact form in two different limits:
\beq
\begin{cases}
{\rm for}\,\,\,  \sqrt{\frac{B}{L}} \ll T \,   \qquad &  z_h=\frac{3}{4 \pi T}  \, , \\
{\rm for}\,\,\, \sqrt{\frac{B}{L}} \gg T  \, , \qquad &   z_h = 6^{1/4} \sqrt{\frac{L}{B}} \, .
\end{cases}
\eeq
The  $T \rightarrow 0$ limit corresponds to the extremal limit:
in this case the near horizon metric is described by and AdS$_2$ $\times R^2$ metric.

 The magnetization of the holographic dual system to the magnetic
 RN black brane in eq. (\ref{Reissner}) 
 is \cite{Hartnoll:2007ai,Hartnoll:2007ih,Ammon:2015wua}:
\beq\label{magrn}
M_{RS}=-\frac{\p \hat{f}_{RS}}{\p B} =-\frac{z_h B}{8 \pi G_N} \, ,
\eeq
where $ \hat{f}_{RS}$ is the free energy density of the magnetic RS black brane.
Evaluation of (\ref{magrn}) gives:
\beq
\begin{cases}
{\rm for}\,\,\,  \sqrt{\frac{B}{L}} \ll T \,   \qquad &  M_{RS} =-\frac{3}{32 \pi^2 G_N} \frac{B}{T }  \, , \\
{\rm for}\,\,\, \sqrt{\frac{B}{L}} \gg T  \, , \qquad &   M_{RS} =-\frac{6^{1/4}}{8 \pi G_N} \sqrt{L B}   \, .
\end{cases}
\eeq

 \section{The vortex lattice}
  \label{sect:CCM}

In this section we  dedicate our study to the range of applied magnetic fields 
in type II superconductors  between the lower critical magnetic field $H_{c1}$, where the magnetic field start
 to penetrate inside the superconductor, and the upper critical field $H_{c2}$, 
 where superconductivity is completely destroyed. In this  regime
 the superconductor is pierced by a lattice of flux vortices.

The numerical solution of the backreacted  holographic vortex
 lattice is in general a rather hard problem, because there is no
cylindrical symmetry and one should solve a system of partial differential
equations involving gravity in three dimensions.

In order to determine the critical magnetic fields and the
magnetization curve in the GL model, 
it turns out that the CCM gives an
excellent approximation \cite{brandt2,brandt2-bis,brandt3}.
In this section we apply the CCM  to the 
an holographic vortex lattice with backreaction. \newline

\subsection{Metric ansatz}

We denote by $R$ the radius of the circular cell in the boundary.
The vortex cell area then is: $A_{cell}=\pi R^2$. 
In our units $R$ is dimensionless, since all scales in the model are set by $L$ and $\kappa$.
We consider the following ansatz for the cylindrical symmetric metric in the bulk:
\bea
\label{metric}
ds^2 &=& \frac{L^2}{y^2}\left\{ -Q_1 y_+^2(1-y^3)dt^2+\frac{Q_2}{1-y^3}dy^2
 \right. \nn\\
&& \left.  +y_+^2Q_4 \left(R dx+\frac{R x}{(1+ R x)^2}y^2Q_3 dy \right)^2
+  y_+^2 Q_5 R^2 x^2 d\theta^2 \right\} \, ,
\eea
where $Q_{1,2,3,4,5}$ are function of the AdS normal coordinate $y$
and of the vortex radial coordinate $x$.
The boundary radial cylindrical coordinate is $r= R \, x $,
in such a way that  $x \in [0,1]$.  
This is very similar to the ansatz
 introduced in \cite{Dias:2013bwa} for the single vortex case.
 The main difference is that the $\hat{x}$ coordinate used in \cite{Dias:2013bwa}
for the single vortex case is different from the $x$ coordinate used here. 
The relation is as follows:
\beq
x=\frac{1}{R} \frac{\hat{x}}{1-\hat{x}} \, , \qquad  \hat{x} \in [0,1] \, .
\eeq

Here $y \in [0,1]$, with $y=0$ being the conformal boundary and $y=1$ 
the black hole horizon.  Without loss of generality,   we can set at the horizon $Q_1(x,1) = Q_2(x,1)$.
 With this choice of conventions, the Hawking temperature is:
\be
\label{T2}
T= \frac{3 y_+}{4\pi}.
\ee

 Note that the metric  (\ref{metric}) is not in the Fefferman-Graham form, i.e.
\beq
ds_{FG}^2=\frac{L^2}{z^2} dz^2 + \gamma_{M N} dw^M dw^M   \, ,
\label{fefferman}
\eeq
where the capital latin letter denote the boundary coordinates $w^M=(t, \tx,  \theta)$
and $y \approx y_+ z$. The FG metric is useful to extract the expectation value of boundary quantities; 
 the form  (\ref{metric}) instead is more convenient to numerically solve
the equations of motion (\ref{eqs}).
The change of variables which brings the metric  (\ref{metric})  in FG form can be expanded as follows
nearby the boundary
\beq
y=y_+ z + \sum_{i=2}^\infty a_i(\tx) z^i \, , 
\qquad x= \tx + \sum_{i=1}^\infty b_i(\tx) z^i \, ,
\label{ee-ab}
\eeq
where some of the lowest order coefficients $(a_i,b_i)$ can be found in Appendix A
of \cite{Tallarita:2019czh}.

The matter field profiles are taken as follows  \cite{Dias:2013bwa}:
\be
\phi = y e^{i n \theta}\left(\frac{Rx}{1+Rx}\right)^n Q_6(x,y), \qquad A_\theta 
= L\left(\frac{Rx}{1+Rx}\right)^2 Q_7(x,y).
\ee
For the purposes of this paper we are only interested in 
winding $n=1$ vortices, therefore we restrict to this value from here on.
 We will impose the boundary condition $\partial_y Q_7(x,0)=0$
in order to describe a superconductor and not a superfluid,
see eqs. (\ref{Atheta}) and (\ref{ee-ab}). 

The equations of motion (\ref{eqs}) lead 
 to a complicated set of  coupled non-linear 
 partial differential equations for the $Q_i(x,y)$ 
which must be solved subject to specific boundary conditions which we discuss below.

The equations cannot be solved by standard numerical procedures in their current form. 
To make the equations elliptic, the DeTurck method is adopted.
 This is explained in detail in \cite{Dias:2015nua} and we will not review it here. 
 For our case the reference metric for the DeTurck procedure is chosen 
 to be the same line element as in eq.(\ref{metric}) with
\be
Q_1 = Q_4=Q_5=1\, , \quad Q_3=0 \, ,
\, \qquad
Q_2=1-\tilde{\alpha} y(1-y) \, ,
\label{alphatilde}
\ee
where the  DeTurck parameter is fixed to
 $\tilde{\alpha}=16 \kappa/9$, in order
 to avoid logarithmic tails in the
 near boundary  Fefferman-Graham expansion of the metric. 
 As a final note, it is crucial that the solutions found must 
 satisfy the vanishing of the DeTurck vector $\xi^a\xi_a =0$, 
 otherwise they are known as Ricci solitons. We have checked 
that this is the case for all the numerical solutions presented
in this paper.

\subsection{Boundary conditions and solutions}

The main idea of the CCM is to use a cylindrical symmetry ansatz
(see figure \ref{fig0})
to approximate an triangular or square Wigner-Seitz cell.
In the GL model, the following conditions are imposed in order to reproduce
the physical vortex lattice \cite{ihle}:
\begin{itemize}
\item the flux of the magnetic field in each cell is the same
as the flux of an elementary vortex
\item  the radial derivative of the scalar condensate
and of the magnetic field vanishes at the boundary of the cell.
\end{itemize}

The coordinate $x=1$ corresponds to the boundary of our cell,
 therefore this is where we have to pay special attention to this boundary condition. 
The geometry of the cell array has to match together in a smooth way.
This gives the following boundary conditions for the metric:
\beq
\partial_x Q_1(1,y)  =  \partial_x Q_2(1,y)=\partial_x Q_4(1,y)=\partial_x Q_5(1,y)=0 \, ,
\qquad
Q_3(1,y)=0 \, .
\label{bou1}
\eeq
The boundary conditions  the gauge fields 
come from the physical requirement that the total magnetic
flux inside a cell is $1/q$, 
which from Stokes' theorem is equivalent to $A_\theta(x=1)=1/q$, i.e.
\beq
 Q_7(1,y) =\frac{(1+R)^2}{qLR^2} \, ,
 \label{bou-QQ7}
\eeq
Eq. (\ref{bou-QQ7}) is also equivalent to the condition that the current 
\beq
J_\mu=i q (\phi^\dagger D_\mu \phi- \phi D_\mu \phi^\dagger) 
\eeq
vanishes at the boundary of a cell.
The circular cell method 
implies that we have to impose the condition $D_x \phi=0$, i.e.
\beq
  \partial_x Q_6(1,y) = - \frac{1}{1+R} Q_6(1,y)\, .
  \label{bou2}
\eeq
Note that the original boundary condition for the circular cell method involved 
setting the derivative of the magnetic field to zero at the boundary of the cell. 
The condition  $\p_x (F^{\mu \nu} F_{\mu \nu})=0$ at the boundary of the cell
follows from eqs. (\ref{bou1}, \ref{bou-QQ7}). \newline

The other boundary conditions are chosen as in  \cite{Dias:2013bwa}:
\begin{itemize}
\item {$y=0$} 
At the conformal boundary we require the metric  to tend to the black brane solution, therefore
\be
Q_1 = Q_2 = Q_4 = Q_5 = 1, \quad  Q_3=0.
\ee
The boundary condition on the scalar field is the Robin condition previously mentioned: 
\be
\partial_y Q_6(x,0) = \frac{\kappa_1}{y_+} Q_6(x,0),
\ee
where $\kappa_1$ is related to $\kappa$ and to the  $\tilde{\alpha}$
parameter as follows:
\beq
\kappa_1=\frac{\tilde{\a} y_+}{4} \, .
\eeq
The boundary condition on the gauge field 
is the one which corresponds to the holographic dual of a superconductor, i.e.  $\partial_y Q_7(x,0)=0$.
\item { $x=0$}
These conditions in the vortex core are derived in the appendix of \cite{Dias:2013bwa},
\bea\nn
\partial_x Q_1(0,y)  &=&  \partial_x Q_2(0,y)=\partial_x Q_4(0,y)=\partial_x Q_5(0,y)=0  , \quad  Q_4(0,y) =  Q_5(0,y),
\nl\nn
\partial_x Q_3(0,y) &=& 2R\; Q_3(0,y), \quad  \partial_x Q_6(0,y) =R\; Q_6(0,y) \quad \partial_x Q_7(0,y) =2R\; Q_7(0,y).
\eea
\item{ $y=1$} 
At the horizon, the only condition that one must satisfy is that $Q_1(x,1) = Q_2(x,1)$.
\end{itemize}

\begin{figure}[ptb]
\begin{subfigure}{.5\textwidth}
\centering
\includegraphics[width=0.9\linewidth]{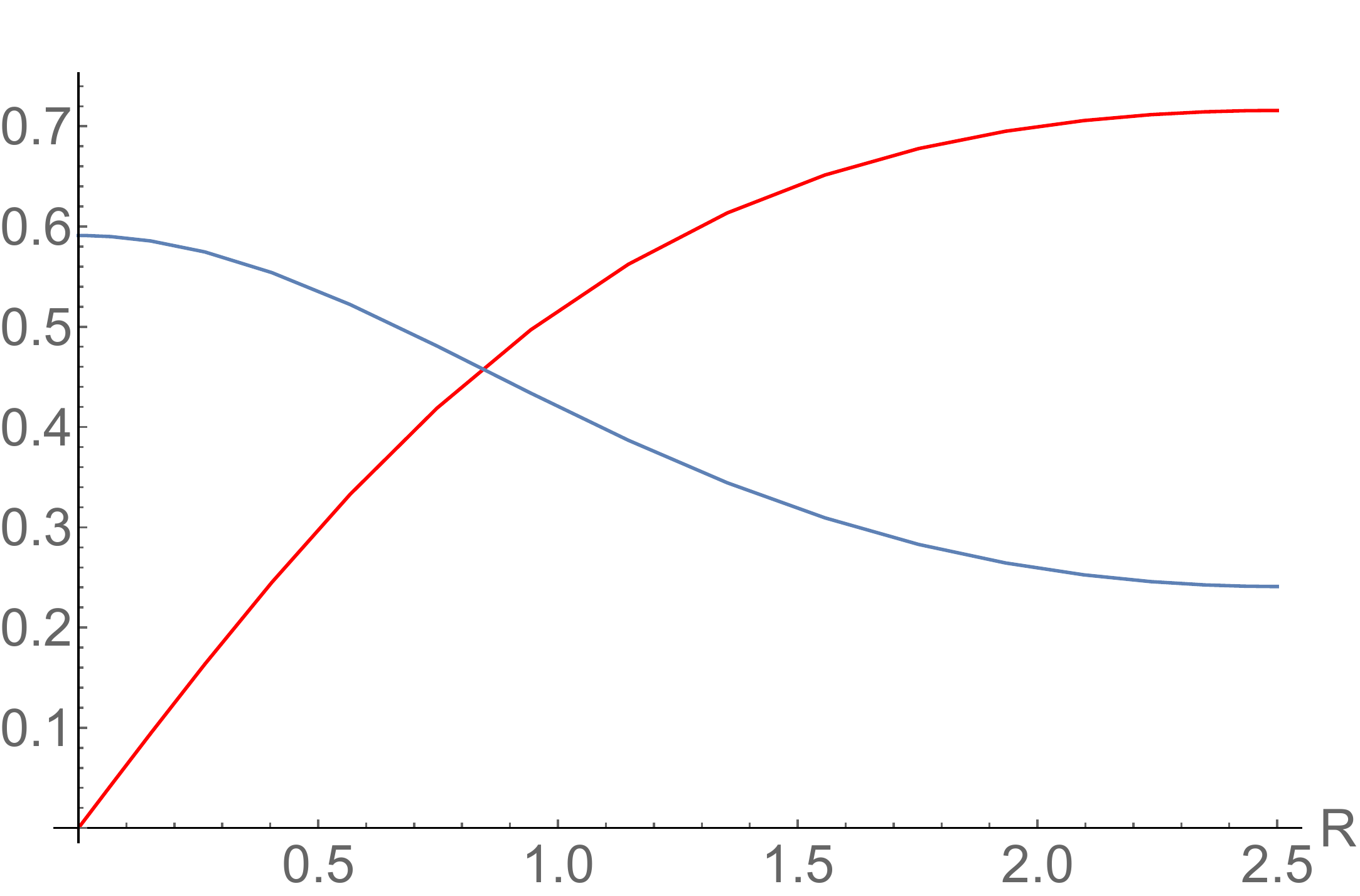}
\caption{$B$ (blue) and scalar field profiles}
\end{subfigure}
\begin{subfigure}{.5\textwidth}
\centering
\includegraphics[width=0.9\linewidth]{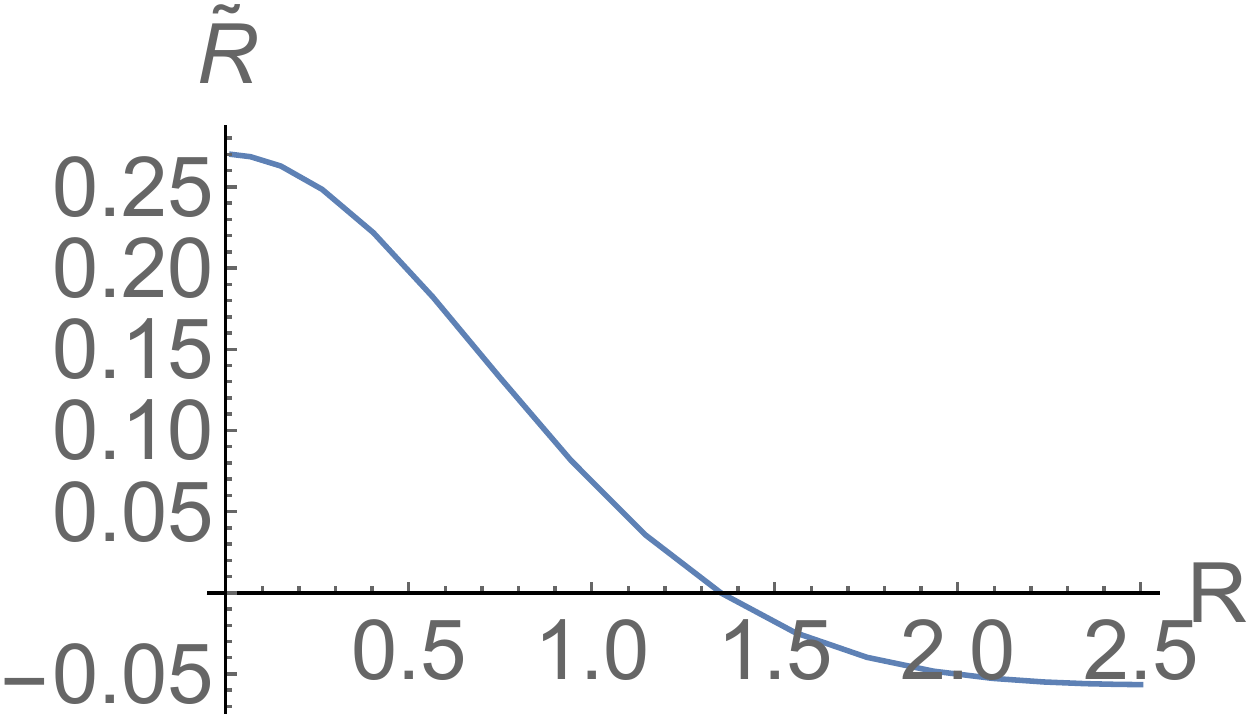}
\caption{induced Ricci scalar at the horizon}
\end{subfigure}
\caption{Some representative solutions of the superconductor case for $R=2.5$ and $q=1$, $\kappa=-1$, $y_+=1$, $G_N=1$, $L=1$. 
The solutions show the field profiles (scalar field in red and magnetic field in blue on the left, Ricci scalar on the right) in one cell of size $R$.}
\label{fig1}
\end{figure}

Some representative solutions are shown in figure \ref{fig1}, showing the profiles for the magnetic field, 
the scalar field and the induced horizon Ricci scalar in the vortex cell. \newline

We have also checked that the induced Ricci scalar $\tilde{R}$ on the horizon
 is smooth ($\partial_x \tilde{R}=0$) at $x=1$, and that there are therefore no gravitational 
 singularities at the borders of the cell. Before proceeding to calculate 
 the magnetization in the full magnetic field regime, we first make, 
 in the next subsection, an important check on the accuracy of the circular cell approximation.

\subsection{The critical magnetic field limit $H_{2c}$}
 \label{sect:critical}

In the limit where the magnetic field approaches from below the upper critical field $H_{2c}$, 
the scalar condensate is small and the authors of \cite{Maeda:2009vf} derived
 an analytic solution for the holographic vortex lattice for a non-backreacting model. 
 This solution was based on a different holographic model, namely one containing 
 a chemical potential and no quartic potential.
  However, since the solution is separable in the bulk and spatial coordinates, 
 we can use the spatial part of the solution as the spatial part of the solution to our model in this limit. 
 This fact allows us to compare the full analytic result to our circular cell approximations taken close to the upper critical magnetic field limit.
 
 The main idea is that nearby the critical magnetic field $H_{2c}$
 the AdS bulk equations are separable. The spatial part of the scalar 
 condensate equation then reproduces the profiles found by Abrikosov
\cite{Abrikosov:1956sx}.

This solution is of the form 
\be
\phi(x_1,y_2, y) = \frac{\rho_0(y)}{L}\sum_{l=-\infty}^{l=\infty}c_l e^{i p_l x_2}\gamma_0(x_1 ;p_l) \, ,
\ee
where $(x_1,x_2)$ denote the two spatial directions and
\be
\gamma_0(x_1 ;p_l) = \exp\left(-\frac{1}{2}\left(\frac{x_1}{r_0}-p r_0\right)^2\right) \, ,
\qquad
p_l=\frac{2\pi l}{a_1 r_0} \, , 
\qquad c_l = \exp \left(-i \frac{\pi a_2}{a_1^2}l^2\right) \, , 
\ee
for constant $p$ and $r_0$. $\rho_0(u)$ here denotes the profile of the solution on the $AdS$ bulk direction.  
The parameters $a_i$ control the geometry of the lattice solution. For square lattices, we simply have $a_1 = a_2$, while for
triangular lattices the relation is
\be
\frac{a_2}{a_1} = \frac{a_1}{2} = 3^{-1/4} \sqrt{\pi}.
\ee

 In other words, for our model $\rho_0(y)$ will be different, but the rest of the solution is the same. 
 The comparison of the spatial profiles at the boundary is independent of $\rho_0(y)$.\newline

\begin{figure}[ptb]
\begin{subfigure}{.5\textwidth}
\centering
\includegraphics[width=0.9\linewidth]{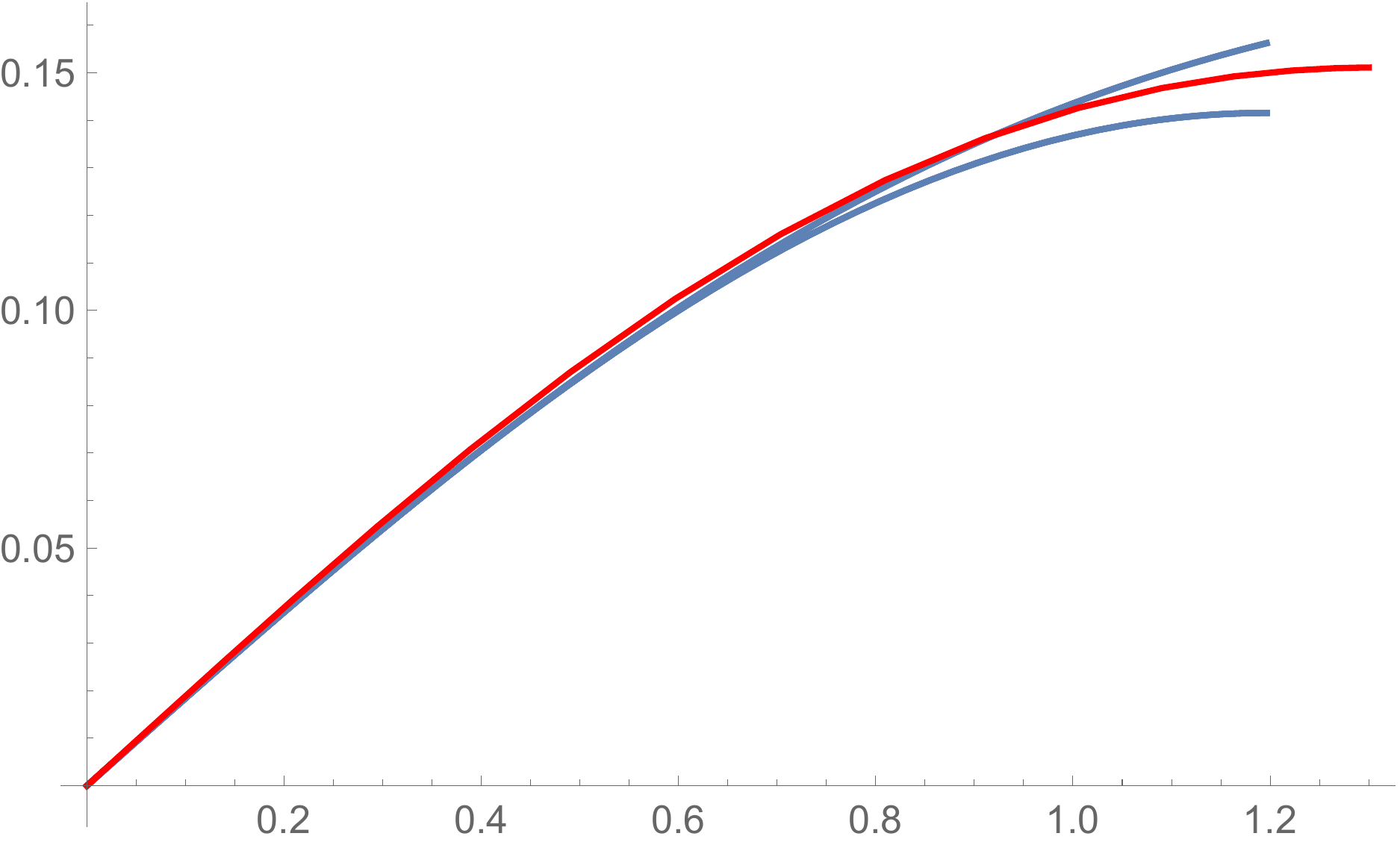}
\caption{square}
\end{subfigure}
\begin{subfigure}{.5\textwidth}
\centering
\includegraphics[width=0.9\linewidth]{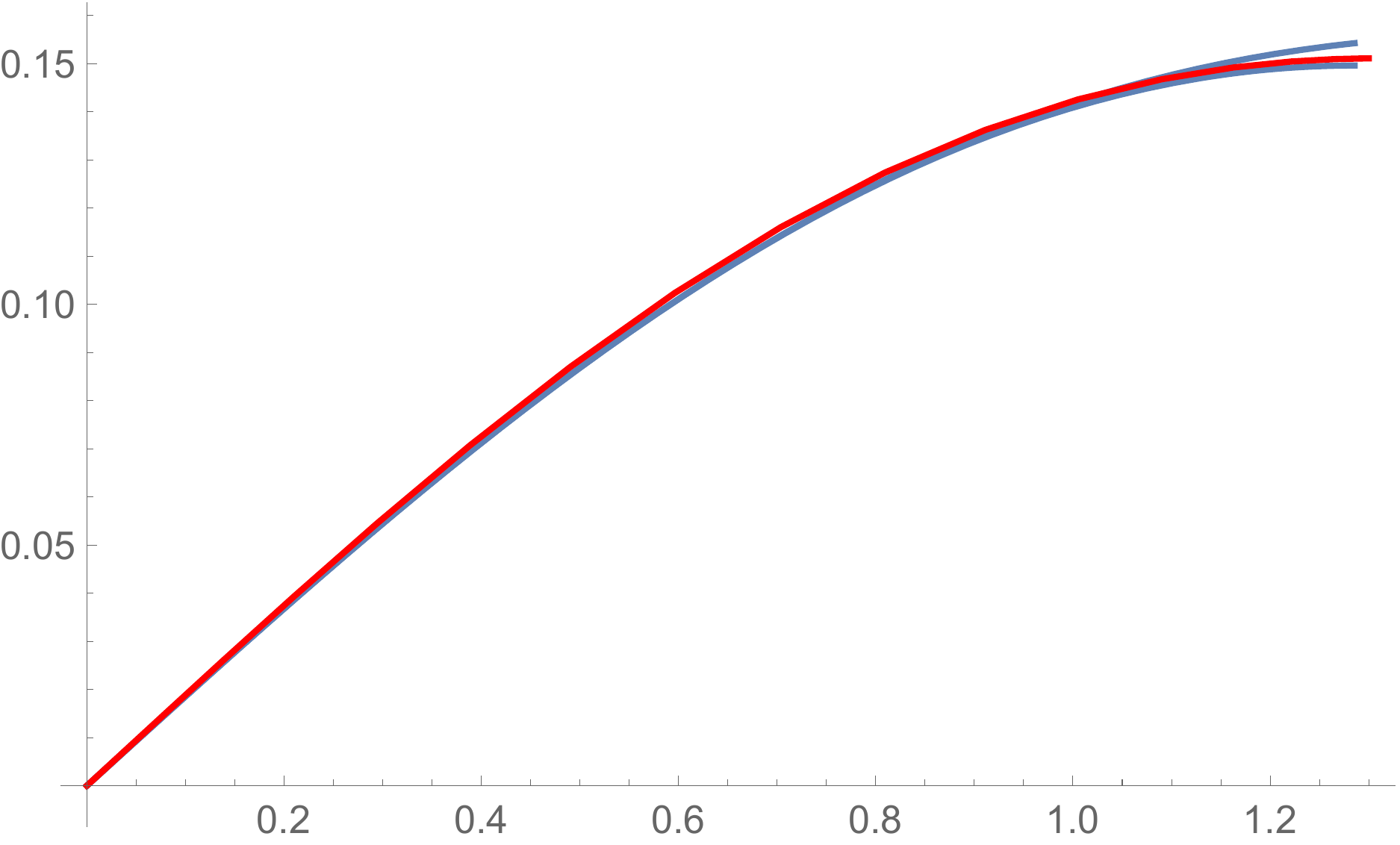}
\caption{triangular}
\end{subfigure}
\caption{Comparison of circular cell method with square and triangular analytic lattice for $H_{c2}- B \approx 0.01$. 
The plot shows the scalar field profile $\phi$. Red line in between the other blue lines is the circular cell result. 
The other two correspond to the analytic solution in the $x_1$ (lower)  and $x_2$ (higher) directions.}
\label{fig2}
\end{figure}

With this solution in hand, we performed an explicit check of the accuracy 
of our circular approximation by comparing vortex profiles in both $x_1$ and $x_2$
 directions in this limit of our solution. 
 There is a remarkable agreement between the two, see figure \ref{fig2}, especially with the triangular case.

\section{Free energy and Magnetization}

The free energy of the cell  is defined by
\be\label{free}
F = E -TS,
\ee
where $E$ is the energy of our solution, $T$ is the temperature and $S$ is its entropy. 

The energy can be computed  by performing holographic renormalization
  in FG coordinates, see \cite{Dias:2013bwa,Tallarita:2019czh} for details.
  The result for $E$ is:
\be
E = \frac{-y_+^2}{G}R^2\int_0^1 x dx \left(\frac{3}{48}y_+
 Q_1^{(3)}(x)+\frac{17 y_+\alpha + 160 \kappa_1}{256}\left(\frac{R x}{1+R x}\right)^2 Q_6(x)^2\right) \, ,
\ee
where $Q_i^{(k)}$ are the  series expansion of the functions  $Q_i$
in powers of $y$ around $y=0$:
\beq
Q_i= \sum_{k=0}^\infty Q_i^{(k)} y^k \, .
\label{ee}
\eeq
The entropy can be computed from the area of the horizon,
see \cite{Dias:2013bwa}:
\be
S = \frac{\pi}{2}y_+^2R^2\int_0^1 x dx \sqrt{Q_4(x, 1)Q_5(x,1)}.
\ee

The final expression for the averaged free energy density is therefore
\be
\bar{f}= 2\pi F/(\pi R^2) = \frac{2(E-TS)}{R^2}
\ee
where we included angular factors in the integration and divided by the cell area $\pi R^2$. 
We denote by $\bar{B}$ the averaged magnetic field in the cell:
\be
\bar{B}=\frac{1}{q} \frac{2}{ R^2} \, .
\ee

With these results in hand we can now proceed to calculate  the magnetization for the holographic superconductor vortex array. 
This is the main advantage of using the circular cell method, as the calculation for the magnetization would otherwise involve solving the full three dimensional vortex array system, or resorting to magnetic field limits in which the system is tractable analytically. With this method, which as we saw is extremely accurate at least for the case of magnetic fields close to the upper critical value, we can extend the calculation to the whole range of magnetic field values. 

Our superconducting boundary condition on $Q_7$ means the dual current $J_\phi =0$
and corresponds to infinite boundary gauge coupling $g \rightarrow \infty$.
The finite value of the magnetization  $M$, which is proportional to $g$ and $J_\phi$
then arises from a zero times infinity limit which is difficult to compute directly;
it is more straightforward to use the definition of magnetization
using free energy.

\begin{figure}[ptb]
\centering
\includegraphics[width=0.7\linewidth]{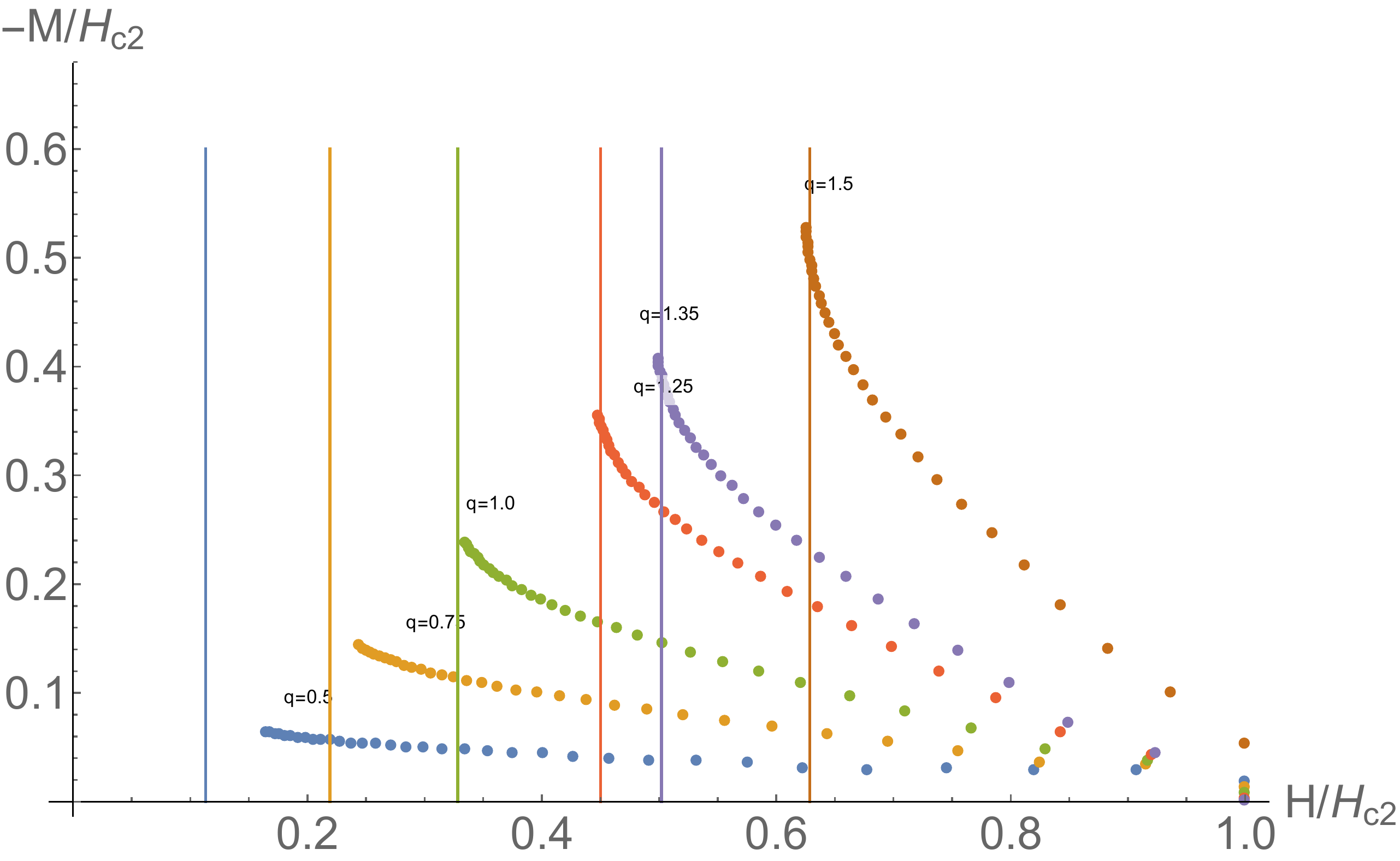}
\caption{Magnetization plot for several values of $q$   including values of $H_{c1}$ (vertical lines).
We set $y_+=1$,  $\kappa=-1$, $L=1$, $G_N=1$.
}
\label{fig22}
\end{figure}

The applied magnetic field can be defined as the source to which the
magnetic field is coupled and it can be extracted from the derivative of the 
free energy density with respect to $\bar{B}$:
\be
 H = \frac{\partial \bar{f}}{\partial \bar{B}} = - q \frac{R^3}{4}\frac{\partial \bar{f} }{\partial R} \, .
\ee 
This should be compared with a region of space where no 
superconductor is present, which is described by the Reissner-Nordstrom (RS) solution. 
We denote by $H_{RS}$ the applied magnetic
field of the RS black brane with the same temperature and 
magnetic field $B=\bar{B}$:
\be
 H_{RS} = \frac{\partial \hat{f}_{RS}}{\partial B}= -M_{RS} \, ,
\ee 
see eq. (\ref{magrn}).
The magnetization  can be expressed as the difference between
the RS magnetic field and the applied magnetic field in 
the presence of the superconductor:
\be
M= H_{RS}- H \, .
\ee
Numerical plots of this results, for varying values of $q$, are shown in figure \ref{fig22}.
These values are all chosen inside the type II parameter space.
The type II/type I transition is close to $q L =2$.

We can calculate the value of $H_{c1}$ for different values of $q$ from 
the single vortex limit. Using the free energy functional for the single vortex, 
we have that \cite{Domenech:2010nf},
\be
H_{c1}= \frac{q}{2\pi}(F_1-F_0),
\ee
where $F_n$ (defined in equation \ref{free}) denotes the free energy of the state with $n$ vortices.
 The results of these calculations are shown as the vertical lines (one for each $q$) in figure \ref{fig22}
 and they are in good agreement with the magnetization curve.

 \section{Discussion}
 \label{sect:conclu}
 
 In this paper we studied holographic vortex lattices in a holographic
 superconductor model in asymptotically AdS$_4$ spacetime,
 using the circular cell approximation.
We computed the magnetization curves and we found
a qualitative agreement with the ones computed for
a Ginzburg-Landau superconductor (see e.g. \cite{brandt3}). 
In the limit nearby the upper critical magnetic field, 
we showed that the circular cell method gives a good approximation of
 the Abrikosov solution.

In general, there are two limits in which the backreaction of
the scalar field on the metric is small:
\begin{itemize}
\item The upper critical magnetic field regime, i.e. $H \approx H_{2c} $  with $H< H_{2c}$.
We checked  that the spatial profiles found by Abrikosov are reproduced (see section \ref{sect:critical}).
\item The $T \approx T_c$ limit, where $T_c$ is the critical temperature for the condensation of $\phi $ in eq. (\ref{Tcritica}).
Above $T_c$ the scalar field $\phi$ is zero, and so  the condensate
is small approaching this temperature from below. In this limit we expect that the Landau-Ginzburg theory
provides a good effective description of the physics, including the vortex lattice.
\end{itemize}

In the numerical example discussed in figure \ref{fig22} 
we use the numerical values $y_+=1$,  $\kappa=-1$.
From eqs. (\ref{T2}) and  (\ref{Tcritica}), these  values correspond to 
$T \approx 0.24$ and $T_c \approx 0.62$,  and so to  $T_c/T \approx 2.6$.
This should be  quite far away from the small field limit.
 Still we get magnetization curves rather similar to the ones computed in
 \cite{brandt3} for a  Ginzburg-Landau superconductor.
 
In the regime $ T \ll T_c$ (or equivalently $ T \ll |\kappa|  $), we expect that the backreaction of the scalar field $\phi$
 is increasingly important. Unfortunately, our numerical calculations 
 become challenging in this regime. It would be interesting to further study 
 this limit to check if some interesting behaviour appears in the magnetization curve.

A more accurate numerical study using
a square and triangular lattice ansatz is desiderable. 
In the case of the GL superconductor, the triangular
lattice is energetically preferred, with a smaller (at the per cent level)
energy per unit of magnetic flux. It would be interesting
to check if the triangular lattice is preferred also 
for holographic superconductors. Moreover, this numerical study
would allow to compare the distribution of the magnetic field
and the magnetization for each lattice symmetry
and to compute the flux-line lattice elastic shear modulus.

Another promising direction is to extend these studies to the 
case of non-abelian vortex strings
\cite{Hanany:2003hp,Auzzi:2003fs,Shifman:2004dr,Shifman:2012vv}.
The force between two non-BPS non-abelian vortex
 strings\footnote{In the BPS case there is no net force between vortices
 with arbitrary orientation.
A large vortex moduli space,
which includes relative positions and  internal orientations,
 appears \cite{Hashimoto:2005hi,Eto:2005yh,Auzzi:2005gr}.}
  depends on the relative orientation of the internal degrees
of freedom localised on the vortex \cite{Auzzi:2007iv,Auzzi:2007wj}.
A rich structure of vortex lattice phase transitions may be realisable
in these case. Vortex lattices in a weakly coupled model of non-abelian
vortices were studied in \cite{Tallarita:2017opp}.
It would be interesting to perform similar studies for holographic 
non-abelian vortices \cite{Tallarita:2015mca,Tallarita:2019czh}. 

 \section*{Acknowledgments}

G.T. is funded by a Fondecyt grant number 11160010.

\end{document}